\documentstyle[12pt]{article}
\begin{document}

\title{\hfill{\normalsize DO-TH-99-21 } \\ \hfill{\normalsize OHSTPY-HEP-T-99-026} \\
\vspace{0.5cm}
 Searching for Rephase-Invariant CP- and CPT-violating Observables 
in Meson Decays }
\author{ K.C. Chou$^{a}$,\  W.F. Palmer$^{b}$,\  E.A. Paschos$^{c}$ 
and  Y. L.  Wu$^{a,c}$ \\ 
\\
$a:$\  Institute of Theoretical Physics, Chinese Academy of Sciences 
\\ Beijing 100080, China \\ 
\\
 $b:$\  Department of Physics, Ohio-State University 
 \\ Columbus, OH 43210, USA \\ 
 \\
 $c:$\  Institut f\"{u}r Physik, Universit\"{a}t Dortmund 
 \\ D-44221 Dortmund, Germany 
 }
\date{palmer@mps.ohio-state.edu \\ paschos@hal1.physik.uni-dortmund.de \\  
ylwu@itp.ac.cn}
\maketitle
\begin{abstract} 
 We present a general model-independent and rephase-invariant formalism 
that cleanly relates CP and CPT noninvariant observables to 
the fundamental parameters.  Different types of CP and CPT violations
 in the $K^{0}$-, $B^{0}$-, $B_{s}^{0}$- and $D^0$-systems are explicitly defined. 
Their importance for interpreting experimental measurements of CP and CPT violations 
is emphasized. In particular, we show that the time-dependent measurements allow 
one to extract a clean signature of CPT violation. 
\end{abstract}
%\narrowtex

\newpage
 
 \section{Introduction}
 
For the discrete symmetries of nature, violations have been observed for C, P
and the combined CP symmetries\cite{P,P1,P2,P3,CP}. In fact two types of CP
violation have now been established in the $K$-meson system. It remains an
active problem of research to observe CP asymmetries in heavier mesons. In
addition there is  new interest in investigations of properties
of the CPT symmetry\cite{CPTT}. Up to now, there are only bounds on
CPT-violating parameters\cite{CPT}, which are sensitive to the magnitude of
amplitudes, but tests of the relative phases have not yet been carried 
out.
 
   In this article we present tests of CPT and CP, separately, and discuss
which measurements  distinguish between the various symmetry breaking terms. In
addition, we derive formulae  which are manifestly invariant under rephasing of
the original mesonic states.  The hope is to call attention to several 
measurements which will be accessible to experiments in the future. 

Our paper is organized as follows:  In section 2, we present a complete set of
parameters characterizing CP, T and CPT nonconservation arising from the mass 
matrix, i.e., the so-called indirect CP-, T- and CPT-violation.  A set of direct 
CP-, T- and CPT-violating parameters originating from the decay amplitudes  are 
defined in section 3. In section 4, we defind all possible independent observables
and relate them directly to fundamental parameters which are manifestly rephasing
invariant  and can be applied to all meson decays. The various types of CP and
CPT violation are classified,  indicating how one can extract purely CPT or CP
violating effects.  In section 5, we investigate in detail the time evolution
of mesonic decays and  introduce several time-dependent CP- and CPT-asymmetries
which  allow one to measure separately the indirect CPT- and CP-violating
observables as well as  direct CPT- and CP-violating observables. In
particular, we show how one can extract a clean  signature of CPT violation
from asymmetries in neutral meson decays.  In section 6, we apply the general
formalism to the semileptonic and nonleptonic  K-meson decays and show how many
rephasing invariant CP and CPT observables can be extracted  separately. Our
conclusions are presented in the last  section.

  \section{CP- and CPT-violating Parameters in Mass Matrix} 

  Let $M^0$ be the neutral meson (which can be $K^0$ or $D^0$ or $B^0$ or $B^{0}_{s}$) 
and $\bar{M}^0$ its antiparticle. The evolution of $M^0$ and $\bar{M^0}$ states is dictated by
\begin{equation}
\frac{d}{dt} \left( \begin{array}{c} 
M^{0} \\ \bar{M}^{0} \end{array} \right) 
= -i \left( \begin{array}{cc}
H_{11} & H_{12} \\
H_{21} & H_{22} \\
\end{array} \right) 
\left( \begin{array}{c} 
M^{0} \\ \bar{M}^{0} \end{array}\right)
\end{equation}
with $H_{ij} = M_{ij} -i \Gamma_{ij}/2$ the matrix elements, and $M_{ij}$, $\Gamma_{ij}$ being 
the dispersive and absorptive parts, respectively.

 The eigenvalues of the Hamiltonian are 
 \begin{eqnarray}
 H_{1} & = &  H_{11} - \sqrt{H_{12}H_{21}}\ \frac{1-\Delta_{M}}{1+\Delta_{M}}\ , \nonumber  \\ 
H_{2} & = &  H_{22} + \sqrt{H_{12}H_{21}}\ \frac{1-\Delta_{M}}{1+\Delta_{M}}\ ,  
\end{eqnarray}
with
\begin{equation}
\frac{1-\Delta_{M}}{1+\Delta_{M}} = \left[ 1 + \frac{\delta_{M}^{2}}{2} -
\delta_{M} \sqrt{1 + \frac{\delta_{M}^{2}}{4}} \right]^{1/2} \ , \qquad and 
\qquad \delta_{M} = \frac{H_{22} - H_{11}}{\sqrt{H_{12}H_{21}}}
 \end{equation}
We note already that $\delta_M$ is invariant under rephasing of the states 
$M^0$ and $\bar{M^0}$. The eigenfunctions of the Hamiltonian define the physical states. 
Following Bell and Steinberger\cite{BS}, $M^0$ and $\bar{M^0}$ mix with each other 
and form two physical mass eigenstates
\begin{equation} 
M_1  =  p_{S}| M^0 > + q_{S} | \bar{M^0} >, \qquad 
 M_2  =  p_{L}| M^0 > - q_{L} | \bar{M^0} >
\end{equation}
with normalization $|p_{S}|^{2} + |q_{S}|^{2} = |p_{L}|^{2} + |q_{L}|^{2} = 1 $.
The coeficients are given by
\begin{eqnarray}
\frac{q_{S}}{p_{S}} & = & \frac{q}{p}\ \frac{1+\Delta_{M}}{1-\Delta_{M}} 
\equiv \frac{1-\epsilon_{S}}{1+ \epsilon_{S}} \ ,  \qquad
\frac{q_{L}}{p_{L}} = \frac{q}{p}\ \frac{1-\Delta_{M}}{1+\Delta_{M}} 
\equiv \frac{1-\epsilon_{L}}{1+ \epsilon_{L}} \nonumber \\
\frac{q}{p} & = & \sqrt{\frac{H_{21}}{H_{12}}} \equiv 
\frac{1-\epsilon_{M}}{1+ \epsilon_{M}} 
\end{eqnarray} 
We have also introduced the paramters  $\epsilon_{S,L,M}$
 following ref.\cite{CRONIN}. In the CPT conserving case they reduce to the known parameter
$\epsilon_{M}$. Thus we have a complete description of the physical states in terms of 
the mass matrix, and the time evolution is determined by the eigenvalues: 
\begin{equation}
H_{1} = M_{1} - i\Gamma_{1}/2; \qquad H_{2} = M_{2} - i\Gamma_{2}/2
\end{equation}
and is given simply by
\begin{equation}
M_{1} \rightarrow e^{-iH_{1}t} M_{1}; \qquad M_{2} \rightarrow e^{-iH_{2}t} M_{2}
\end{equation}

 We discuss next several properties related to the symmetries of the system. The parameters
 $\delta_M$ and $|q/p|$ are rephasing invariant and so are also other parameters defined 
 in terms of them. CPT invariance requires $M_{11} = M_{22}$ and $\Gamma_{11} = \Gamma_{22}$, 
 and implies that $\delta_M = 0$. Thus the difference between $q_{S}/p_{S}$ and $q_{L}/p_{L}$
 represents a signal of CPT violation. In other words, $\Delta_{M}$ different from zero 
 indicates CPT violation.
 
   CP invariance requires the dispersive and absorptive parts of $H_{12}$ and
$H_{21}$ to be,  respectively, equal and implies $ q/p  = 1$. Also if T
invariance holds, then  independently of CPT symmetry, the dispersive and
absorptive  parts of $H_{12}$ and $H_{21}$ must be equal up to a total relative
common phase, implying $ |q/p|  = 1$. Therefore a  $Re\epsilon_M$
different from zero describes CP and  T nonconservation and can be present even
when CPT is conserved. Finally,  two parameters, $\epsilon_{M}$
describing CP violation  with T nonconservation and $\Delta_{M}$ characterizing
CPT violation with CP nonconservation, are related to $\epsilon_{S}$ and
$\epsilon_{L}$ via 
\begin{equation}
\epsilon_{S} = \frac{\epsilon_{M} - \Delta_{M}}{1 -\epsilon_{M}\Delta_{M}}; \qquad 
\epsilon_{L} = \frac{\epsilon_{M} + \Delta_{M}}{1 +\epsilon_{M}\Delta_{M}}
\end{equation}
and reduce to those given in \cite{CRONIN} when neglecting the quadratic 
term $\epsilon_{M} \Delta_{M}$.  This is a complete set of parameters describing CP, T and CPT
nonconservation which originates in the mass matrix (indirect). In the next section we discuss 
additional parameters originating in the decay amplitudes (direct) as well as from the mixing 
between mass matrix and decay amplitudes (mixed-induced).

 \section{CP- and CPT-violating Parameters in Decay Amplitudes }

 Let $H_{eff}$ be the effective Hamiltonian which contains CPT-even 
 $H_{eff}^{(+)}$ and CPT-odd $H_{eff}^{(-)}$ parts, i.e., 
 \begin{equation}
  H_{eff} = H_{eff}^{(+)} + H_{eff}^{(-)}
  \end{equation}
 with
 \begin{equation}
 (CPT) H_{eff}^{(\pm)} (CPT)^{-1} = \pm\ H_{eff}^{(\pm)}
 \end{equation}  
Let $f$ denote the final state of the decay and $\bar{f}$ 
its charge conjugate state. The decay amplitudes of $M^0$ 
are defined as
\begin{eqnarray}
g & \equiv & <f|H_{eff}| M^0>=\sum_{i} (A_{i} + B_{i}) e^{i\delta_{i}} \equiv 
\sum_{i} (|A_{i}|e^{i\phi_{i}^{A}} + |B_{i}|e^{i\phi_{i}^{B}}) e^{i\delta_{i}} , \nonumber \\
\bar{h}  & \equiv & <\bar{f}|H_{eff}|M^0 > = \sum_{i} (C_{i} + D_{i}) 
e^{i\delta_{i}} \equiv \sum_{i} (|C_{i}|e^{i\phi_{i}^{C}} + |D_{i}|e^{i\phi_{i}^{D}}) 
e^{i\delta_{i}}  
\end{eqnarray}
with $A_{i}$ and $C_{i}$ being CPT-conserving amplitudes
\begin{equation}
<f|H_{eff}^{(+)}| M^0> \equiv \sum_{i} A_{i} e^{i\delta_{i}}\ , \qquad 
 <\bar{f}|H_{eff}^{(+)}| M^0> \equiv \sum_{i} C_{i} e^{i\delta_{i}}
\end{equation}
and $B_{i}$ and $D_{i}$ being CPT-violating amplitudes
\begin{equation}
<f|H_{eff}^{(-)}| M^0> \equiv \sum_{i} B_{i} e^{i\delta_{i}}\ , \qquad 
 <\bar{f}|H_{eff}^{(-)}| M^0> \equiv \sum_{i} D_{i} e^{i\delta_{i}} \ .
\end{equation}
Here we have used the notation of ref.\cite{BARMIN} for the amplitude $g$, and
have introduced a new amplitude $\bar{h}$. The second amplitude is absent when one considers 
only K-meson decays and neglects possible violation of $\Delta S = \Delta Q$ rule as was the case 
in ref.\cite{BARMIN}. This is because the K-meson decays obey $\Delta S = \Delta Q$ 
rule via weak interactions of the standard model. The reason is simple
since the strange quark can only decay to the up quark. 
In the case of $B$-, $B_s$- and $D$-meson systems both amplitudes $g$ and $\bar{h}$ exist
via the $W$-boson exchange of weak interactions since both $b$-quark and $c$-quark will 
have two different transitions due to CKM quark mixings, i.e., $b\rightarrow c, u$ and 
$c\rightarrow s, d$ (for explicit decay modes see the classification for the processes 
given in section 5). $\phi_{i}^{I}$ ($I=A,B,C,D$) are weak phases and $\delta_{i}$ are 
strong phases from final state interactions. The subscrpts $i=1,2, \cdots$ denote 
various strong interacting final states, such as the different isospin states. 
For CP transformation, we adopt the phase convention 
\begin{equation} 
CP | M^0> =  |\bar{M^0}>\  , \qquad CP |\bar{M^0}> =  | M^0> \ ,
\end{equation}
It is then not difficult to show that the decay amplitudes of the charge conjugate meson 
$\bar{M^0}$ have the following form    
\begin{eqnarray}
\bar{g}  & \equiv & <\bar{f}|H_{eff}|\bar{M^0}> = \sum_{i} (A_{i}^{\ast} - B_{i}^{\ast}) 
e^{i\delta_{i}}  \equiv 
\sum_{i} (|A_{i}|e^{-i\phi_{i}^{A}} - |B_{i}|e^{-i\phi_{i}^{B}}) e^{i\delta_{i}}, \nonumber \\
h & \equiv & <f|H_{eff}|\bar{M^0} > = \sum_{i} (C_{i}^{\ast} - D_{i}^{\ast}) 
e^{i\delta_{i}} \equiv 
\sum_{i} (|C_{i}|e^{-i\phi_{i}^{C}} - |D_{i}|e^{-i\phi_{i}^{D}}) e^{i\delta_{i}}\ . 
\end{eqnarray}

  In analogy to the indirect CP- and CPT-violating parameters $\epsilon_{S,L,M}$ from mass
matrix, we define now parameters containing direct CP and CPT violations 
\begin{equation}
\varepsilon_{M}'  \equiv  \frac{1-h/g}{1+h/g}, \ 
 \bar{\varepsilon}_{M}' \equiv \frac{1-\bar{g}/\bar{h}}{1+ \bar{g}/\bar{h}}; \ 
 \varepsilon_{M}''  \equiv  \frac{1-\bar{g}/g}{1+ \bar{g}/g}, \   
\bar{\varepsilon}_{M}'' \equiv \frac{1- h/\bar{h}}{1+ h/\bar{h}}
\end{equation}
For final states which are CP conjugate, i.e., $ |\bar{f}> = CP | f > = |f>$,
the relations $h = \bar{g}$ and $ \bar{h} = g$ hold, and thus the four parameters
are reduced to two independent ones: $\varepsilon_{M}' = \varepsilon_{M}''$ and
$\bar{\varepsilon}_{M}' = \bar{\varepsilon}_{M}''$. 

  The symmetry properties of the amplitudes are as follows. If CP is conserved, 
independently of CPT symmetry, one has $\bar{g}/g = 1$ and $ h/\bar{h}= 1$, which implies
\[ A_i = A_{i}^{\ast} \ , \qquad C_i = C_{i}^{\ast}\ , \qquad 
B_i = -B_{i}^{\ast}\ , \qquad D_i = - D_{i}^{\ast} \]
in other words: 
\[ \phi_{i}^{A}= \phi_{i}^{C} = 0\ , \qquad  \phi_{i}^{B} = \phi_{i}^{D} = \pi/2 \  ,\] 
namely, $A_i$ and $C_i$ are real, while $B_i$ and $D_i$ are imaginary. 

Similarly T invariance exchanges the initial and final states and implies,
independently of CPT symmetry, 
\[ A_i = A_{i}^{\ast} \ , \qquad C_i = C_{i}^{\ast}\ , \qquad 
B_i = B_{i}^{\ast}\ , \qquad D_i =  D_{i}^{\ast} \]
or 
\[ \phi_{i}^{A}= \phi_{i}^{C} = 0\ , \qquad  \phi_{i}^{B} = \phi_{i}^{D} = 0 \ , \] 
namely, all the amplitudes must be real. Finally, conservation of CPT requires 
$B_i = 0$ and $D_i= 0$.  We summarize the results for the amplitudes in Table 1.

{\bf Table 1.}
\begin{center}
\begin{tabular}{|c|c|c|} \hline
 &  CPT-conservation & CPT-Violation  \\  \hline
CP-conservation  &  $A_i = A_{i}^{\ast} \quad C_i = C_{i}^{\ast}$ & 
$B_i = -B_{i}^{\ast} \quad D_i = - D_{i}^{\ast} \quad$ imply T-violation \\ 
 & & \\
T-conservation   &  $A_i = A_{i}^{\ast} \quad C_i = C_{i}^{\ast}$ & 
$B_i = B_{i}^{\ast} \quad D_i =  D_{i}^{\ast} \quad$ imply CP-violation \\ 
& CP $\&$ T conservation &   \\   \hline
\end{tabular}
\end{center}
 
Reading across the first row of the table we have the
conditions for CP conservation, with T conservation (first column) and
without T-conservation (second column). The relations $B_i = -B_{i}^{\ast}$
and $D_i = - D_{i}^{\ast}$ imply T-violation in the presence of CP
conservation. The second row of the table gives the conditions when T is
conserved, with CP conservation (first column) or without CP conservation
(second column).  This is a complete set of amplitude with
the $C_i$ and $D_i$ amplitudes introduced for the first time here. As a
consequence, two more CP- and CPT-violating parameters $\varepsilon_M$ and
$\bar{\varepsilon}_M$ in eq. (16) are needed. 

 In summary of this section, we have the following conclusions.
{\it Values for Re$\varepsilon_{M}''$ and Re$\bar{\varepsilon}_{M}''$
different from zero describe CP nonconservation independently of T and CPT
symmetries. The presence of $B_i's$ and $D_i's$ indicate simultaneous
nonconservation of:  CPT and either of CP or T. Zero $\varepsilon_{M}''$ 
and $\bar{\varepsilon}_{M}''$ with nonzero $Im\varepsilon_{M}'$ and 
$Im\bar{\varepsilon}_{M}'$ implies T nonconservation. Finally, 
zero $B_i$ and $D_i$, and complex $A_i$ and $C_i$ signal CPT conservation 
with CP and T violations.} Note that the latter case is more difficult 
to establish experimentally since it requires the observation of a relative 
phase between two amplitudes distinguished with the help of specific 
quantum numbers. This was the case with the $\epsilon'/\epsilon$ parameter 
in K-meson decays.

\section{Rephase Invariant CP- and CPT-violating Observables }

 The $\varepsilon$-type parameters defined in eqs.(5) and (16) can not be related to 
physical observables since they are not rephasing invariant. Let us introduce 
 CP- and CPT-violating observables by considering the ratio,
\begin{equation} 
\hat{\eta}_{f} \equiv \frac{q_{S}}{q_{L}} 
\frac{<f |H_{eff} | M_{2} >}{< f |H_{eff} | M_{1} >}  
= \frac{q_{S}}{q_{L}}\frac{p_{L}}{p_{S}}\frac{1 - r_{f}^{L}}{1 + r_{f}^{S}}
\end{equation}
which enters to the time evolution of the decay amplitudes (see eqs. 27 and 28). 
The parameters $q_{S,L}$ and $p_{S,L}$ were defined in section 2, and we also 
introduce the notation

\[ r_{f}^{S} = (q_{S}/p_{S})(h/g) \]
with a similar definition for $r_{f}^{L}$. Note that the factor $q_S/q_L$ is necessary 
for the normalization and also rephase invariance, which has not been always included in the 
literature. In the CPT-conserving case \cite{PW} this factor is equal to 
unity. One can simply see from the definitions in eqs.(3)-(5) that $\hat{\eta}_{f}$ is rephasing 
invariant.  The factor $q_{S}p_{L}/ p_{S} q_{L} = (1+\Delta_{M})^2/(1-\Delta_{M})^2$ is 
rephase-invariant since $\Delta_M$ has this property. The ratios
$r_{f}^{L,S} = (q_{L,S}/p_{L,S})(h/g)$ are also rephase-invariant. To see that, let us make a phase 
redefinition $|M^0> \rightarrow e^{i\phi} |M^0>$, then $|\bar{M}^0> \rightarrow e^{-i\phi} |\bar{M}^0>$, 
$H_{12} \rightarrow e^{-2i\phi} H_{12}$ and $H_{21} \rightarrow e^{2i\phi} H_{21}$, 
as well as $h \rightarrow e^{-i\phi} h$ and $g \rightarrow e^{i\phi} g$, 
thus $(q_{S}/ p_{S}, q_{L}/ p_{L}) \rightarrow 
e^{2i\phi} (q_{S}/ p_{S}, q_{L}/ p_{L})$ and $h/g \rightarrow 
e^{-2i\phi} h/g$, which makes $r_{f}^{L,S} = (q_{L,S}/p_{L,S})(h/g)$ 
manifestly rephase-invariant. 

  It is seen that the rephase-invariant quantities $r_{f}^{L,S}$ and
$\hat{\eta}_{f}$ are given by the product of complex parameters arising from
the mass mixing $(q_{L,S}/p_{L,S})$ and from amplitudes $(h/g)$. To separately
define the rephase-invariant CP- and CPT-violating observables originating from
the mass mixing and from the amplitudes, some algebra is neccesary\footnote{The 
algebra is described in ref.\cite{PW}}, but
it is not difficult to show that $\hat{\eta}_{f}$ can be rewritten as
\begin{equation} 
\hat{\eta}_{f} \equiv \frac{1}{1-\eta_{\Delta}} 
\left[ \eta_{\Delta} + 
\frac{a_{\epsilon_{S}} + \hat{a}_{\epsilon'} + i\ \hat{a}_{\epsilon_{S} + 
\epsilon'}}{2 + a_{\epsilon_{S}} \hat{a}_{\epsilon'} + 
\hat{a}_{\epsilon_{S} \epsilon'}} \right]
\end{equation}
where we have used the definitions
\begin{eqnarray}
a_{\epsilon_{S}} & = & \frac{1-|\frac{q_{S}}{p_{S}}|^{2}}{1+|\frac{q_{S}}{p_{S}}|^{2}} 
= \frac{2Re \epsilon_{S}}{1 + |\epsilon_{S}|^{2}} = 
\frac{a_{\epsilon} - a_{\Delta}}{1- a_{\epsilon}a_{\Delta}}, \nonumber \\
a_{\epsilon_{L}} & = & \frac{1-|\frac{q_{L}}{p_{L}}|^{2}}{1+|\frac{q_{L}}{p_{L}}|^{2}} 
= \frac{2Re \epsilon_{L}}{1 + |\epsilon_{L}|^{2}} = 
\frac{a_{\epsilon} + a_{\Delta}}{1+ a_{\epsilon}a_{\Delta}} \\
\eta_{\Delta} & = & \frac{2\Delta_{M}}{1+ \Delta_{M}^{2}} = \frac{a_{\Delta} + 
i a'_{\Delta}\sqrt{1 - a_{\Delta}^{2} - a_{\Delta}^{'2}} }{1- a_{\Delta}^{'2}}  \nonumber
\end{eqnarray}
with 
\begin{eqnarray} 
a_{\epsilon} & = &  \frac{1 - |q/p|^2}{1 + |q/p|^2} = 
\frac{2 Re \epsilon_{M}}{1 + |\epsilon_{M}|^2} \ ,   \\
a_{\Delta} & = & \frac{2Re \Delta_{M}}{1 + |\Delta_{M}|^{2}}, \qquad 
a'_{\Delta} = \frac{2 Im \Delta_{M}}{1 + |\Delta_{M}|^{2}} \nonumber 
\end{eqnarray}
The definitions of $\hat{a}_{\epsilon'}$, $\hat{a}_{\epsilon_{S} + \epsilon'}$ and 
$\hat{a}_{\epsilon_{S} \epsilon'}$ are given in the appendix. 
The reader should note that quantities without a hat contain either only CP or only CPT 
nonconserving effects, and with a hat contain both CP- and CPT-nonconserving effects.

   As $a_{\epsilon}$, $\hat{a}_{\epsilon'}$, $\hat{a}_{\epsilon + \epsilon'}$
and $\hat{a}_{\epsilon \epsilon'}$ (for their definitions see appendix) are all
rephase-invariant, so are also $\hat{a}_{\epsilon_{S} + \epsilon'}$ and
$\hat{a}_{\epsilon_{S} \epsilon'}$. Note that only three of them are
independent since  $(1- a_{\epsilon}^{2})(1 - \hat{a}_{\epsilon'}^{2}) =
\hat{a}_{\epsilon + \epsilon'}^{2} + (1+ \hat{a}_{\epsilon \epsilon'})^{2}$.
Another rephase-invariant direct CP and CPT noninvariant observable is defined
as \begin{equation} 
\hat{a}_{\epsilon''} = \frac{1-|\bar{g}/g|^{2}}{1 + |\bar{g}/g|^2 } 
= \frac{2 Re \varepsilon_{M}'' }{1 + |\varepsilon_{M}''|^2 } = \frac{a_{\epsilon''} 
+ a_{\varepsilon\Delta} + a'_{\Delta\Delta}}{1 + 
a'_{\varepsilon\Delta} + a_{\Delta\Delta} }
\end{equation} 
where the definitions for $a_{\epsilon''}$ , $a_{\varepsilon\Delta}$, $a'_{\Delta\Delta}$,  
$a'_{\varepsilon\Delta}$ and $a_{\Delta\Delta}$ are presented in the appendix.
 Analogously, one has 
\begin{equation}
\hat{\eta}_{\bar{f}} \equiv \frac{q_{S}}{q_{L}}
\frac{<\bar{f} |H_{eff} | M_{2} > }{<\bar{f} |H_{eff} | M_{1} > } = 
\frac{1}{1- \eta_{\Delta}} 
\left[ \eta_{\Delta} + \frac{a_{\epsilon_{S}} + \hat{a}_{\bar{\varepsilon}'} 
+ i\ \hat{a}_{\epsilon_{S} + \bar{\varepsilon}'}}{2 + a_{\epsilon_{S}} 
\hat{a}_{\bar{\varepsilon}'} + \hat{a}_{\epsilon_{S} \bar{\varepsilon}'}} \right]
\end{equation}
and
\begin{equation}
\hat{a}_{\bar{\epsilon}''} = \frac{1-|\bar{h}/h|^{2}}{1 + |\bar{h}/h|^2 } 
= \frac{2 Re \bar{\varepsilon}_{M}'' }{1 + |\bar{\varepsilon}_{M}''|^2 } 
= \frac{a_{\bar{\epsilon}''} 
+ a_{\bar{\varepsilon}\bar{\Delta}} + a'_{\bar{\Delta}\bar{\Delta}} }{1 + 
a'_{\bar{\varepsilon}\bar{\Delta}} + a_{\bar{\Delta}\bar{\Delta}} }
\end{equation} 
with $\bar{\Delta}_{i} = D_{i}/C_{i}$. 

 One of the interesting cases occurs when the final states are CP eigenstates, 
i.e., $f^{CP} = f$, and in this case $h =\bar{g}$ (or $C=A$ and $D=B$). 
As a consequence, we find 
\begin{eqnarray}
& & \hat{a}_{\epsilon'} = \hat{a}_{\epsilon''}, \qquad 
a_{\epsilon'} = a_{\epsilon''} \nonumber \\
& & \hat{a}_{\epsilon + \epsilon'} = \frac{1}{1 + 
a'_{\varepsilon\Delta} + a_{\Delta\Delta} } [ a_{\epsilon + \epsilon'} 
+ a_{\epsilon + \epsilon'_{\Delta}} + a_{\epsilon + \epsilon'_{\Delta\Delta} } ]
\end{eqnarray}
where the explicit definitions for $a_{\epsilon + \epsilon'}$, 
$a_{\epsilon + \epsilon'_{\Delta}}$ and $a_{\epsilon + \epsilon'_{\Delta\Delta} }$ 
are again given in the appendix.

 To see explicitly how many rephase invariant CPT and CP observables may be
separately measured from  experiments, let us consider the case for which
the final states are CP eigenstates and suppose that the violations are small 
so that one could only keep the linear terms of the rephase invariant CPT- and 
CP-violating observables. With this consideration, the observable
$\hat{\eta}_f$ is simplified  \begin{equation}  \hat{\eta}_f \simeq
\frac{1}{2}[ a_{\epsilon} + a_{\epsilon'} + a_{\Delta} + a_{\epsilon\Delta}  +
a'_{\Delta\Delta} + i( a_{\epsilon + \epsilon'} + a'_{\Delta} + a_{\epsilon +
\epsilon'_{\Delta\Delta}}  + a_{\epsilon + \epsilon'_{\Delta}} ) ] 
\end{equation} where the definitions for all the rephase invariant quantities
are given in the appendix.  Those with index $\Delta$ are the CPT-violating
observables, the others are CP-violating ones which have been discussed in
ref.\cite{PW}. 

  The formalism so far involves many equations which include CP and CPT violation effects either separately 
or mixed together. It has several advantages in comparison with other articles\cite{LL,BARMIN}:
\begin{enumerate}
\item The formalism is more general than the ones reported in the literature and can be applied 
not only to the K-meson decays but also all other heavier meson decays. 
\item All observables are manifestly rephasing invariant and well defined by directly relating 
     to the hadronic mixing matrix elements and decay amplitudes of mesons.    
\item All possible independent observables are classified, which enables one to separately 
      measure different types of CPT- and CP-violating observables and to extract purely CPT or CP 
      violation effects. 
\item The formalism is more elegantly designed for extracting various rephase invariant CPT- and 
    CP- violating observables from time-dependent measurements of meson decays, which will be discussed 
    in detail in the next section.    
\end{enumerate}

   We have thus defined all possible rephase-invariant CP and CPT noninvariant 
observables in terms of eight parameters related to CP and CPT breaking quantities arising either from 
mixing or phases of amplitude. The eight parameters are classified as follows: $\epsilon_{M}$ is 
an indirect CP-violating parameter and $\Delta_{M}$ the indirect CPT-violating
parameters; the parameters $\varepsilon''_{M}$ and $\bar{\varepsilon}''_{M}$ will be decomposed 
into four parameters, $\epsilon''_{M}$, $\bar{\epsilon}''_{M}$, $\Delta_{i}$ and 
$\bar{\Delta}_{i}$, where $\epsilon''_{M}$ and $\bar{\epsilon}''_{M}$ define direct 
CP-violating paramters, $\Delta_{i}$ and $\bar{\Delta}_{i}$ describe direct CPT-violating 
parameters. $\varepsilon'_{M}$ and $\bar{\varepsilon}'_{M}$ contain the ratio of 
the two decay amplitudes and can be associated with direct CP and CPT violation, 
as well as the interference between indirect and direct CP and CPT violations. 
All the CP and CPT violations can be well defined and in general classified into 
the following types: 
\begin{enumerate}
\item purely indirect CP and CPT violations which are given by 
the rephase-invariant CP-violating observable $a_{\epsilon}$ and CPT-violating 
observables $a_{\Delta}$ and $a'_{\Delta}$. 
\item purely direct CP and CPT 
violations which are characterized by the rephase-invariant CP-violating 
observables $a_{\epsilon''}$ and $a_{\bar{\epsilon}''}$ and CPT-violating 
observables $a_{\varepsilon\Delta}$, $a'_{\varepsilon\Delta}$, 
$a_{\Delta\Delta}$, $a'_{\Delta\Delta}$,  $a_{\bar{\varepsilon}\bar{\Delta}}$, 
$a'_{\bar{\varepsilon}\bar{\Delta}}$, $a_{\bar{\Delta}\bar{\Delta}}$ and 
$a'_{\bar{\Delta}\bar{\Delta}}$. 
\item Mixed-induced CP and CPT violations which 
are described by CP-violating observables $a_{\epsilon + \epsilon'}$ and   
$a_{\epsilon + \bar{\epsilon}'}$ and CPT-violating observables 
$a_{\epsilon + \epsilon'_{\Delta}}$, $a_{\epsilon + \epsilon'_{\Delta\Delta}}$,
$a_{\epsilon + \bar{\epsilon}'_{\Delta}}$ and
$a_{\epsilon + \bar{\epsilon}'_{\Delta\Delta}}$. 
\end{enumerate}
For the case that the final states are CP eigenstates, one has 
$\hat{a}_{\epsilon'} = \hat{a}_{\epsilon''}= \hat{a}_{\bar{\epsilon}'}
= \hat{a}_{\bar{\epsilon}''}$. Thus, in this case $\hat{a}_{\epsilon'}$ 
and $ \hat{a}_{\bar{\epsilon}'} $ also indicate purely direct CP and CPT 
violations. When the final states are not CP eigenstates, 
$\hat{a}_{\epsilon'}$ and $ \hat{a}_{\bar{\epsilon}'} $ do not,  in general, 
provide a clear signal of direct CP violation although they contain direct 
CP and CPT violations. Their deviation from the values $\hat{a}_{\epsilon'} = \pm 1, \  0$ and 
$ \hat{a}_{\bar{\epsilon}'}= \mp 1, \  0 $ can arise from different CKM angles,  
final state interactions, or different hadronic form factors, 
but not necessarily from CP and CPT violations.

\section{Extraction of CP- and CPT-violating Observables}

 In order to measure the rephase-invariant observables defined above, 
we consider the proper time evolution\cite{LW,PZ} of the neutral mesons 
 \begin{equation}
|M^{0}(t) >  =  \sum_{i=1}^{2} \xi_{i} e^{-i(m_{i} - i \Gamma_{i}/2)t } 
|M_{i} >  \ ; \qquad
|\bar{M^{0}}(t) > =  \sum_{i=1}^{2} \bar{\xi_{i}} e^{-i(m_{i} - 
i \Gamma_{i}/2)t } |M_{i} > 
\end{equation}
with $\xi_{1}= q_{L}/(q_{S}p_{L}+q_{L}p_{S})$ and $\xi_{2}= q_{S}/(q_{S}p_{L}+q_{L}p_{S})$
for a pure $M^0$ state at $t=0$ as well as  $\bar{\xi}_{1}= p_{L}/(q_{S}p_{L}+q_{L}p_{S})$ 
and $\bar{\xi}_{2}= -p_{S}/(q_{S}p_{L}+q_{L}p_{S})$ for a pure $\bar{M}^0$ state at $t=0$. 
Thus the decay amplitudes of $M^{0}$ and $\bar{M}^{0}$ at the time $t$ will 
be given by
 
\begin{eqnarray}
 {\cal A}(t) & = & <f|M^{0}(t)> = \frac{<f|M_{1}>}{p_{S}} \frac{1-\eta_{\Delta}}{2} 
\left( e^{-iH_{1}t} + \hat{\eta}_{f} e^{-iH_{2}t} \right) \ , \\
\bar{{\cal A}}(t) & = & <\bar{f}|\bar{M}^{0}(t)> = \frac{<\bar{f}|M_{1}>}{q_{S}} 
\frac{1-\eta_{\Delta}}{2} \left( \frac{1+\eta_{\Delta}}{1-\eta_{\Delta}}e^{-iH_{1}t} 
-\hat{\eta}_{\bar{f}} e^{-iH_{2}t} \right) 
\end{eqnarray}

It follows now that the time-dependent decay rates are 
\begin{eqnarray}
 & & \Gamma ( M^{0} (t) \rightarrow f ) \propto |{\cal A}(t)|^2 
= (|g|^2 + |h|^2)\frac{2 + a_{\epsilon_{S}}\hat{a}_{\epsilon'}
+ \hat{a}_{\epsilon_{S}\epsilon'}}{1 + a_{\epsilon_{S}}}  e^{-\Gamma t} \nonumber \\ 
& & \cdot \{ 
 [ \frac{1+ a_{\epsilon_{S}} \hat{a}_{\epsilon'}+ 
(a_{\epsilon_{S}} + \hat{a}_{\epsilon'}) Re \eta_{\Delta}  + 
\hat{a}_{\epsilon_{S} + \epsilon'} Im \eta_{\Delta}}{2 + a_{\epsilon_{S}}\hat{a}_{\epsilon'}
+ \hat{a}_{\epsilon_{S}\epsilon'} } -Re\eta_{\Delta} + |\eta_{\Delta}|^{2} ] \cosh (\Delta\Gamma t) 
\nonumber \\ 
 & & + [ \frac{1 + a_{\epsilon_{S} \hat{\epsilon}'}
- (a_{\epsilon_{S}} + \hat{a}_{\epsilon'}) Re \eta_{\Delta} - 
\hat{a}_{\epsilon_{S} + \epsilon'}Im \eta_{\Delta}}{2 + a_{\epsilon_{S}}\hat{a}_{\epsilon'}
+ \hat{a}_{\epsilon_{S}\epsilon'} }
- Re\eta_{\Delta} ] \sinh (\Delta\Gamma t) \\
 & & + [ \frac{(a_{\epsilon_{S}} + \hat{a}_{\epsilon'})(1- Re \eta_{\Delta}) -
\hat{a}_{\epsilon_{S} + \epsilon'}Im \eta_{\Delta}}{2 + a_{\epsilon_{S}}\hat{a}_{\epsilon'}
+ \hat{a}_{\epsilon_{S}\epsilon'}} + Re\eta_{\Delta} - |\eta_{\Delta}|^{2}]
 \cos (\Delta m t) \nonumber \\
 & & + [ \frac{\hat{a}_{\epsilon_{S} + \epsilon'}(1-Re \eta_{\Delta}) + 
(a_{\epsilon_{S}} + \hat{a}_{\epsilon'}) Im \eta_{\Delta}}{2 + a_{\epsilon_{S}}\hat{a}_{\epsilon'}
+ \hat{a}_{\epsilon_{S}\epsilon'}} + Im\eta_{\Delta}  ]\sin (\Delta m t) \} 
\nonumber 
\end{eqnarray}
and 
\begin{eqnarray}
 & & \Gamma ( \bar{M}^{0} (t) \rightarrow \bar{f} ) \propto |\bar{{\cal A}}(t)|^2 
= (|\bar{g}|^2 + |\bar{h}|^2)\frac{2 + a_{\epsilon_{S}}\hat{a}_{\bar{\epsilon}'}
+ \hat{a}_{\epsilon_{S}\bar{\epsilon}'}}{1 + a_{\epsilon_{S}}}  e^{-\Gamma t} \nonumber \\ 
& & \cdot \{ 
 [ \frac{1+ a_{\epsilon_{S}} \hat{a}_{\bar{\epsilon}'}+ 
(a_{\epsilon_{S}} + \hat{a}_{\bar{\epsilon}'}) Re \eta_{\Delta}  + 
\hat{a}_{\epsilon_{S} + \bar{\epsilon}'} Im \eta_{\Delta}}{2 + 
a_{\epsilon_{S}}\hat{a}_{\bar{\epsilon}'}+ \hat{a}_{\epsilon_{S}\bar{\epsilon}'} } 
-Re\eta_{\Delta} - |\eta_{\Delta}|^{2} ] \cosh (\Delta\Gamma t) 
\nonumber \\ 
 & & + [ \frac{1 + a_{\epsilon_{S} \hat{\bar{\epsilon}}'}
- (a_{\epsilon_{S}} + \hat{a}_{\bar{\epsilon}'}) Re \eta_{\Delta} - 
\hat{a}_{\epsilon_{S} + \bar{\epsilon}'}Im \eta_{\Delta}}{2 + 
a_{\epsilon_{S}}\hat{a}_{\bar{\epsilon}'}+ \hat{a}_{\epsilon_{S}\bar{\epsilon}'} }
- Re\eta_{\Delta} ] \sinh (\Delta\Gamma t) \\
 & & - [ \frac{(a_{\epsilon_{S}} + \hat{a}_{\bar{\epsilon}'})(1+ Re \eta_{\Delta}) +
\hat{a}_{\epsilon_{S} + \bar{\epsilon}'}Im \eta_{\Delta}}{2 + 
a_{\epsilon_{S}}\hat{a}_{\bar{\epsilon}'}
+ \hat{a}_{\epsilon_{S}\bar{\epsilon}'}} + Re\eta_{\Delta} + |\eta_{\Delta}|^{2}]
 \cos (\Delta m t) \nonumber \\
 & & -[ \frac{\hat{a}_{\epsilon_{S} + \bar{\epsilon}'}(1 +Re \eta_{\Delta}) - 
(a_{\epsilon_{S}} + \hat{a}_{\bar{\epsilon}'}) Im \eta_{\Delta} }{2 + 
a_{\epsilon_{S}}\hat{a}_{\bar{\epsilon}'}+ \hat{a}_{\epsilon_{S}\bar{\epsilon}'}} 
- Im\eta_{\Delta}  ]\sin (\Delta m t) \} 
\nonumber 
\end{eqnarray}
where $\Delta \Gamma = \Gamma_2 - \Gamma_1 $ and $\Delta m = m_2 - m_1 $. Here we have 
omitted the integrals from the phase space. Similarly, one can easily write down the decay 
rates $\Gamma (M^{0}(t) \rightarrow \bar{f} )$ and 
$\Gamma (\overline{M}^{0}(t) \rightarrow f )$, and then the time-dependent CP 
and CPT  asymmetries are defined by the difference 
between two decay rates. In addition, in studies of the time dependence one can isolate each
of four-terms. One can introduce several asymmetries 
from the decay rates $\Gamma (M^{0}(t) \rightarrow f )$, 
$\Gamma (\overline{M}^{0}(t) \rightarrow \bar{f}  )$ , 
$\Gamma (M^{0}(t) \rightarrow  \bar{f} )$ and $\Gamma (\overline{M}^{0}(t) 
\rightarrow f )$ . Obviously, the time dependences contains a lot of information.
Therefore studies of time evolution can eliminate the various components (hamonics) in 
$\cos(\Delta m t)$, $\sin(\Delta m t)$, $\cosh(\Delta \Gamma t)$ and $\sinh(\Delta \Gamma t)$.
We now proceed to apply the above general analysis to specific processes. 
As in the ref.\cite{PW}, we may classify the processes into four scenarios:  
  
 i) \  $M^{0} \rightarrow f $ ($M^{0} \not\rightarrow \bar{f} $) , \ 
$\overline{M}^0 \rightarrow \bar{f}$ ($\overline{M}^0 \not \rightarrow f$) 
, this is the case when $f$ and $\bar{f}$ are not a common final 
state of $M^{0}$ and $\overline{M}^{0}$. 
Examples are:\  $M^0 \rightarrow M'^- \bar{l} \nu $, $\bar{M}^{0} 
\rightarrow M'^+ l \bar{\nu} $; \  $B^0 \rightarrow D^- D_{s}^{+} $, \ 
$D^- K^+$, $\pi^- D_{s}^{+}$, $\pi^-  K^+$, \  
$\bar{B}^{0} \rightarrow D^+ D_{s}^{-} $,
$D^+ K^-$, $\pi^+ D_{s}^{-}$, $\pi^+ K^-$;\  
 $B_{s}^{0} \rightarrow D_{s}^{-} \pi^+$, $D_{s}^{-} D^{+}$, $K^{-} \pi^{+}$
, $K^{-} D^+$, \  $\overline{B}^{0}_{s} \rightarrow 
D_{s}^{+} \pi^-$, $D_{s}^{+} D^-$, $ K^{+} \pi^{-}$, $K^+ D^-$. 
This scenario also applies to charged meson decays.

 ii) \  $M^{0} \rightarrow (f = \bar{f}, \  f^{CP} =  f) \leftarrow 
 \overline{M}^0$, this is the decay to a common final state which is CP eigenstate. 
Such as $B^{0} (\bar{B}^{0} )$, $D^{0} (\bar{D}^{0} )$, 
$K^{0} (\bar{K}^{0} )$  $\rightarrow \pi^{+} \pi^{-} $, $\pi^{0} \pi^{0}$, 
\ $\cdots $. For the final states such as $\pi^- \rho^+$ and $\pi^+ \rho^-$ 
, although each of them is not a CP eigenstate of $B^{0}(\bar{B}^{0})$ 
or $D^0 (\bar{D}^0)$, one can always decompose them into CP eigenstates 
as $(\pi \rho)_{\pm} = (\pi^- \rho^+ \pm \pi^+ \rho^- )$ with 
$CP (\pi \rho)_{\pm} = \pm (\pi \rho)_{\pm}$. This reconstruction is 
meaningful since $\pi^- \rho^+$ and  $\pi^+ \rho^-$ have the same weak 
phase as they contain the same quark content.  

 iii) \  $M^{0} \rightarrow ( f, \  f \not \rightarrow f^{CP})  \leftarrow 
 \overline{M}^0$, i.e., the final states are common final states but  
are not charge conjugate states. For example, 
$B^{0} (\bar{B}^{0} ) \rightarrow K_{S} J/\psi $, 
$B^{0}_{s} (\bar{B}^{0}_{s} ) \rightarrow K_{S} \phi $ and 
$D^{0} (\bar{D}^{0} ) \rightarrow K_{S} \pi^{0} $, $K_{S}\rho^{0}$. 

 iv)  \  $M^{0} \rightarrow (f \  \& \  \bar{f}, \  f^{CP} \neq f ) \leftarrow 
 \overline{M}^0$ , i.e., both $f$ and $\bar{f}$ are the common final 
states of $M^0$ and $\overline{M}^0$, but they are not CP eigenstates. 
This is the most general case. For example, \  
$B^{0} (\bar{B}^{0} ) \rightarrow D^- \pi^{+}$, $\pi^{-} D^{+}$ ; \    
$D^{-} \rho^{+}$, $\rho^{-} D^{+}$;\  
$B^{0}_{s} (\bar{B}^{0}_{s} ) \rightarrow D_{s}^{-} K^{+}$,  $K^{-} D_{s}^{+}$;\ 
 $D^0 (\bar{D}^{0}) \rightarrow K^- \pi^{+} $, $K^+ \pi^{-}$.

 In this paper, we will only elaborate on the first two scenarios. In scenario i), 
the amplitudes $h$ and $\bar{h}$ are zero, thus $\hat{a}_{\epsilon'} = 
- \hat{a}_{\bar{\epsilon}'} = 1$, $ \hat{a}_{\epsilon + \epsilon'} = 0  = 
\hat{a}_{\epsilon + \bar{\epsilon}'} $ and 
$\hat{a}_{\epsilon \epsilon'} = -1 = \hat{a}_{\epsilon \bar{\epsilon}'}$. For this case, 
the time-dependent rates of eqs.(29) and (30) will become very simple, 
\begin{eqnarray}
 \Gamma( M^{0} (t) \rightarrow f ) & \propto & |{\cal A}(t)|^2 
=|g|^2  e^{-\Gamma t}\cdot \{ (1 + |\eta_{\Delta}|^{2}) \cosh \Delta\Gamma t 
\nonumber \\
& - & 2Re\eta_{\Delta}\sinh\Delta\Gamma t 
 + (1 - |\eta_{\Delta}|^{2}) \cos \Delta m t + Im\eta_{\Delta} \sin \Delta m t  \} 
\nonumber \\
 \Gamma( \bar{M}^{0} (t) \rightarrow \bar{f} ) & \propto & |\bar{{\cal A}}(t)|^2 
=|\bar{g}|^2  e^{-\Gamma t}\cdot \{ (1 + |\eta_{\Delta}|^{2}) \cosh \Delta\Gamma t 
\\ 
& + & 2Re\eta_{\Delta}\sinh\Delta\Gamma t 
 + (1 - |\eta_{\Delta}|^{2}) \cos \Delta m t - Im\eta_{\Delta} \sin \Delta m t \}
\nonumber 
\end{eqnarray}
It is not difficult to show that the other two time-dependent decay rates which are not allowed
at $t=0$, can happen at a later $t$, because the $M^{0}$ develops an $\bar{M}^{0}$ component through 
mixing. They can be simply expressed as
\begin{eqnarray}
 \Gamma( M^{0} (t) \rightarrow \bar{f} ) & \propto &
\frac{g^{2} + |\bar{g}|^2}{2}(1- \hat{a}_{\epsilon''}) 
\left(\frac{1-a_{\epsilon_{S}}}{1 + a_{\epsilon_{S}}}\right)
\frac{(1-a_{\Delta})^{2}}{1-a_{\Delta}^{'2}} \nonumber \\
& & \cdot e^{-\Gamma t} 
(cosh\Delta\Gamma t - \cos \Delta m t )  \nonumber \\
\Gamma( \bar{M}^{0} (t) \rightarrow f ) & \propto &
\frac{g^{2} + |\bar{g}|^2}{2}(1+ \hat{a}_{\epsilon''}) 
\left(\frac{1+a_{\epsilon_{L}}}{1 - a_{\epsilon_{L}}}\right)
\frac{(1-a_{\Delta})^{2}}{1-a_{\Delta}^{'2}}  \\
& & \cdot e^{-\Gamma t} (cosh\Delta\Gamma t - \cos \Delta m t )  \nonumber 
\end{eqnarray}
With these four decay rates, we can define three asymmetries which have the following 
simple forms when neglecting the quadratic and high order terms of the CP and CPT violating 
parameters (i.e., $a_{\Delta}^{2}$, $a_{\Delta}^{'2}$, $a_{\epsilon}a_{\Delta}^{2}$)

\begin{eqnarray}
A_{CP+CPT}(t) & = & \frac{\Gamma(M^{0}(t) \rightarrow f) - \Gamma 
(\overline{M}^{0} (t) \rightarrow \bar{f}) }{\Gamma(M^{0}(t) \rightarrow f) + \Gamma 
(\overline{M}^{0} (t) \rightarrow \bar{f} )}  \nonumber \\
& \simeq & a_{\epsilon''} + a_{\varepsilon\Delta} + \frac{ -a_{\Delta}\sinh \Delta \Gamma t 
+ a'_{\Delta} \sin \Delta m t }{cosh \Delta\Gamma t  + cos\Delta m t} \\
A'_{CP+CPT} (t) & = & \frac{\Gamma(\overline{M}^{0}(t) \rightarrow f) - 
\Gamma (M^{0}(t) \rightarrow \bar{f} )}{\Gamma(\overline{M}^{0}(t) 
\rightarrow f) + \Gamma (M^{0} (t) \rightarrow \bar{f} )} \nonumber \\
& \simeq & a_{\epsilon''} + a_{\varepsilon\Delta} + 2a_{\epsilon} \\
A''_{CP+CPT}(t) & = & \frac{\Gamma(M^{0}(t) \rightarrow f) - \Gamma (\overline{M}^{0}(t)
\rightarrow f )}{\Gamma(M^{0}(t) \rightarrow f) + 
\Gamma(\overline{M}^{0} (t) \rightarrow f )} \nonumber \\
& \simeq & \frac{\cos\Delta m t - a_{\epsilon} \cosh \Delta \Gamma t 
-a_{\Delta}\sinh \Delta \Gamma t + a'_{\Delta} \sin \Delta m t }{\cosh \Delta \Gamma t - 
a_{\epsilon}\cos\Delta m t -a_{\Delta}\sinh \Delta \Gamma t + a'_{\Delta} \sin \Delta m t}   
\end{eqnarray}
Their exact expressions can be found in the appendix. From the time-dependent measurements 
of the above asymmetries, one shall be able to extract all observables:  
$\Delta m $, $\Delta \Gamma $, $a_{\epsilon}$, $a_{\Delta}$, $a'_{\Delta}$ 
and $\hat{a}_{\epsilon''}$. 

  From the above asymmetries, we easily arrive at the following important observations:
\begin{enumerate}   
\item {\it As long as the experimental measurements show that the asymmetry $A_{CP+CPT}(t)$ is not 
a constant and depends on time, it provides a clean signature of indirect 
CPT violation from mixings}. 
\item For the semileptonic decays $M^{0} \rightarrow 
M^{'-} l\nu$ and also for the decay modes in which the final state interactions are absent,
one has $a_{\epsilon''}=0$, $a'_{\Delta\Delta}=0$, $a'_{\varepsilon\Delta}=0$ and 
$\hat{a}_{\epsilon''} = a_{\varepsilon\Delta}/(1 + a_{\Delta\Delta})$, thus 
nonzero $\hat{a}_{\epsilon''}$ will represent direct CPT violation
from amplitudes. For this case, we come to a strong conclusion that 
{\it once the asymmetry $A_{CP+CPT}(t)$ is not zero, then CPT must be violated}.
\item By combining measurement of the above asymmetries from semileptonic
and nonleptonic decays, it allows one, in principle, to separately measure the
indirect CP-violating observable $a_{\epsilon}$ and the direct CP-violating
observable $a_{\epsilon''}$ as well as the indirect CPT-violating observables
$a_{\Delta}$ and $a'_{\Delta}$, and the direct CPT-violating observable
$a_{\epsilon\Delta}$. 
\end{enumerate}  

  We now  discuss scenario ii) in which $\bar{h} = g$ and $h = \bar{g}$, thus
$a_{\epsilon'} = a_{\epsilon''}= a_{\bar{\epsilon}'}= a_{\bar{\epsilon}''}$ 
and  $a_{\epsilon + \epsilon'} = a_{\epsilon + \bar{\epsilon}'}$.  
When neglecting the quadratic and high order terms and using the relations 
and definitions for the rephase-invariant observables, the
time-dependent asymmetry is simply given by 
\begin{equation}
A_{CP+CPT}(t) \simeq  - (a_{\epsilon}+a_{\Delta}) + e^{-\Delta \Gamma t}[
 (a_{\epsilon}+a_{\Delta} + \hat{a}_{\epsilon'}) \cos (\Delta m t) + 
(a'_{\Delta} + \hat{a}_{\epsilon + \epsilon'}) \sin (\Delta m t) ]
\end{equation}  
(The exact expression is given in the appendix.) 

 From the above time-dependent evolution $A_{CP+CPT}(t)$ 
one is able to extract three physical quantities: one of them is the direct CP 
and CPT noninvariant observable $\hat{a}_{\epsilon'}$ and the other two
are the combinations of CP and CPT noninvariant observables $(a_{\epsilon}+a_{\Delta})$ 
and $(a'_{\Delta} + \hat{a}_{\epsilon + \epsilon'})$. Combining these measurements 
with scenario (i), in which the indirect CP and CPT noninvariant observables $a_{\epsilon}$, 
$a_{\Delta}$ and $a'_{\Delta}$ are expected to be determined, one will be able to extract 
the mixed-induced CP and CPT noninvariant observable $\hat{a}_{\epsilon + \epsilon'}$. 
Thus, studies of scenarios (i) and (ii) allow us to separate the three types of CP and CPT
violations.

\section{CP and CPT Violation in K-meson System}

 The formalism and analyses presented above are general and can be used for all 
 neutral meson systems. As a specific application, we are going to consider the K-meson
 system. From semileptonic decays of $K^{0} \rightarrow \pi^{-} + l^{+} +\nu_l $ 
 and $\bar{K}^{0} \rightarrow \pi^{+} + l^{-} + \bar{\nu_l}$, from eqs.(33) and (34), 
 the time-dependent measurements of the asymmetries lead to
 \begin{eqnarray}
A_{CP+CPT}^{K_{l3}}(t) & = & \frac{\Gamma(K^{0}(t) \rightarrow \pi^{-} l^{+}\nu_l) - \Gamma 
(\overline{K}^{0} (t) \rightarrow \pi^{+} l^{-} 
\bar{\nu_l}) }{\Gamma(K^{0}(t) \rightarrow \pi^{-} l^{+}\nu_l) + \Gamma 
(\overline{K}^{0} (t) \rightarrow \pi^{+} l^{-} \bar{\nu_l})}  \nonumber \\
& \simeq &  a_{\varepsilon\Delta} + \frac{-a_{\Delta}\sinh \Delta \Gamma t + 
a'_{\Delta} \sin \Delta m_{K} t }{cosh \Delta\Gamma t  + cos\Delta m_K t}  \ ,  \\
A_{CP+CPT}^{'K_{l3}}(t) & = & \frac{\Gamma(\overline{K}^{0}(t) \rightarrow \pi^{-} l^{+}\nu_l)
 - \Gamma (K^{0} (t) \rightarrow \pi^{+} l^{-} 
\bar{\nu_l}) }{\Gamma(\overline{K}^{0}(t) \rightarrow \pi^{-} l^{+}\nu_l) + \Gamma 
(K^{0} (t) \rightarrow \pi^{+} l^{-} \bar{\nu_l})}  \nonumber \\
& \simeq &  a_{\varepsilon\Delta} + 2 a_{\epsilon} 
\end{eqnarray}
where the direct CP-violating parameter $a_{\epsilon''}$ is expected to be small as the final
state interactions are electromagnetic. It is then clear that 
{\it non-zero asymmetry $A_{CP+CPT}^{K_{l3}}(t)$ is a clean signature
of CPT violation.} Its time evolution allows us to extract direct CPT-violating observable
$a_{\varepsilon\Delta}$ and indirect CPT-violating observables $a_{\Delta}$ and $a'_{\Delta}$.
The combination of the two asymmetries $A_{CP+CPT}^{'K_{l3}}(t)$ and $A_{CP+CPT}^{K_{l3}}(t)$ 
further helps us to extract indirect CP-violating observable $a_{\epsilon}$. 
 
 In the nonleptonic decays with final states being CP eigenstates, the asymmetry 
 $A_{CP+CPT}(t)$ is given in terms of the observables $\hat{a}_{\epsilon'}$ and 
 $\hat{a}_{\epsilon + \epsilon'}$ which concern both CP and CPT 
violations. In general, it is hard to clearly separate CP violation from CPT violation 
in the decay amplitudes, but it would be of interest to look for possibilities 
of establishing CPT violation arising from the decay amplitudes. For the K-meson system, 
there are two unique decay modes $K^{0} (\bar{K}^{0}) \rightarrow \pi^{+}
\pi^{-}$ and $\pi^{0}\pi^{0}$ which are related via isospin symmetry. Their time-dependent 
asymmetries are given by
\begin{eqnarray}
& & A_{CP+CPT}^{(\pi^{+}\pi^{-})}(t)  \simeq  - (a_{\epsilon}+a_{\Delta})  \\ 
& & +  e^{-\Delta \Gamma t}[
 (a_{\epsilon}+a_{\Delta} + \hat{a}_{\epsilon'}^{(+-)} ) \cos (\Delta m_{K} t) + 
(a'_{\Delta} + \hat{a}_{\epsilon + \epsilon'}^{(+-)} ) \sin (\Delta m_{K} t) ]\ , \nonumber \\
& & A_{CP+CPT}^{(\pi^{0}\pi^{0})}(t)  \simeq  - (a_{\epsilon}+a_{\Delta})  \\ 
& & +  e^{-\Delta \Gamma t}[
 (a_{\epsilon}+a_{\Delta} + \hat{a}_{\epsilon'}^{(00)} ) \cos (\Delta m_{K} t) + 
(a'_{\Delta} + \hat{a}_{\epsilon + \epsilon'}^{(00)} ) \sin (\Delta m_{K} t) ]\ . \nonumber
\end{eqnarray}  
It is seen that since the indirect CP-violating observable $a_{\epsilon}$
and indirect CPT-violating observables $a_{\Delta}$ and $a'_{\Delta}$ can be 
extracted from asymmetries in the semileptonic decays, we then can extract 
the direct CP- and CPT-violating observables $\hat{a}_{\epsilon'}^{(+-)}$ and 
$\hat{a}_{\epsilon'}^{(00)}$ as well as mixed-induced CP- and CPT-violating observables
$\hat{a}_{\epsilon + \epsilon'}^{(+-)}$ and $\hat{a}_{\epsilon + \epsilon'}^{(00)}$. 
We now discuss how to extract pure CPT or CP violation effects by using isospin symmetry.   

  When neglecting high order terms, we have 
 \begin{equation}
\hat{a}_{\epsilon'} \simeq a_{\epsilon'} + a'_{\Delta\Delta} + a_{\varepsilon\Delta}, \qquad 
\hat{a}_{\epsilon + \epsilon'} \simeq  a_{\epsilon + \epsilon'} + 
a_{\epsilon + \epsilon'_{\Delta\Delta}} + a_{\epsilon + \epsilon'_{\Delta}}
\end{equation}
Note that their dependence on the final states are understood. Using the isospin symmetry, we find
\begin{eqnarray}
A^{(+-)} & = & \sqrt{\frac{2}{3}} a_{0} + \sqrt{\frac{1}{3}} a_{2} \nonumber \\
 A^{(00)} & = & \sqrt{\frac{1}{3}} a_{0} - \sqrt{\frac{2}{3}} a_{2} 
\end{eqnarray}
with $A^{(+-)}$ and $A^{(00)}$ the 
amplitudes for the decay modes $K^{0} (\bar{K}^{0}) \rightarrow \pi^{+}\pi^{-}$ and 
$K^{0} (\bar{K}^{0}) \rightarrow \pi^{0}\pi^{0}$ respectively, where $a_{0}$ and $a_{2}$ 
correspond to the isospin $I=0$ and $I=2$ amplitudes. The same decomposition 
holds for $B^{(+-)}$ and $B^{(00)}$ amplitudes\footnote{Note that normalization
of  $A^{(00)}$ is smaller by a factor $\sqrt{2}$ than the usual one ocuuring 
in literature.}. Considering the fact that 
$\omega = |A_{2}|/|A_{0}|\simeq 1/22 << 1$ due to the $\Delta I = 1/2$ rule, we obtain 
\begin{eqnarray}
\hat{a}_{\epsilon'}^{(+-)} & \simeq & a_{\epsilon'} + a'_{\Delta\Delta} 
+ \tilde{a}_{\varepsilon\Delta} + a_{\varepsilon\Delta}^{0}, \nonumber \\
\hat{a}_{\epsilon'}^{(00)} & \simeq & -2a_{\epsilon'} -2 a'_{\Delta\Delta} 
-2 \tilde{a}_{\varepsilon\Delta} + a_{\varepsilon\Delta}^{0},
\end{eqnarray} 
and 
\begin{eqnarray}
\hat{a}_{\epsilon + \epsilon'}^{(+-)} & \simeq & a_{\epsilon +\epsilon'}^{0}  
+ a_{\epsilon +\epsilon'_{\Delta\Delta}}^{0} + \tilde{a}_{\epsilon +\epsilon'}
+ \tilde{a}_{\epsilon +\epsilon'_{\Delta\Delta}} + a_{\epsilon +\epsilon'_{\Delta}}
\nonumber \\
\hat{a}_{\epsilon + \epsilon'}^{(00)} & \simeq & a_{\epsilon +\epsilon'}^{0}  
+ a_{\epsilon +\epsilon'_{\Delta\Delta}}^{0} -2\tilde{a}_{\epsilon +\epsilon'}
-2\tilde{a}_{\epsilon +\epsilon'_{\Delta\Delta}} -2 a_{\epsilon +\epsilon'_{\Delta}}
\end{eqnarray} 
with 
\begin{eqnarray}
& & a_{\varepsilon\Delta}^{0} = 2Re\Delta_{0}= 2Re\left(\frac{B_{0}}{A_{0}}\right), \qquad 
 \tilde{a}_{\varepsilon\Delta} = 2Re[\frac{A_{2}}{A_{0}}(\Delta_{2}-\Delta_{0})]
\cos(\delta_{0}-\delta_{2}) \nonumber \\
& & a_{\epsilon +\epsilon'}^{0} = 2\frac{Im\epsilon_{K}}{1 + |\epsilon_{K}|^{2}}
Re\left(\frac{A_{0}^{\ast}}{A_{0}}\right) + 2\frac{1-|\epsilon_{K}|^{2}}{1 
+ |\epsilon_{K}|^{2}}Im\left(\frac{A_{0}^{\ast}}{A_{0}}\right) \nonumber \\
& & \tilde{a}_{\epsilon +\epsilon'} \simeq 4[\frac{Im\epsilon_{K}}{1 + |\epsilon_{K}|^{2}}
Re\left(\frac{A_{2}^{\ast}}{A_{0}}\right) + \frac{1-|\epsilon_{K}|^{2}}{1 
+ |\epsilon_{K}|^{2}}Im\left(\frac{A_{2}^{\ast}}{A_{0}}\right)]\cos(\delta_{0}-\delta_{2}) \\
& & a_{\epsilon +\epsilon'_{\Delta\Delta}}^{0} = -2\frac{Im\epsilon_{K}}{1 + |\epsilon_{K}|^{2}}
Re\left(\frac{A_{0}^{\ast}}{A_{0}}\Delta_{0}^{\ast 2}\right) - 2\frac{1-|\epsilon_{K}|^{2}}{1 
+ |\epsilon_{K}|^{2}}Im\left(\frac{A_{0}^{\ast}}{A_{0}}\Delta_{0}^{\ast 2}\right) \nonumber \\ 
& & \tilde{a}_{\epsilon +\epsilon'_{\Delta\Delta}} \simeq -4[\frac{Im\epsilon_{K}}{1 + 
|\epsilon_{K}|^{2}}Re\left(\frac{A_{2}^{\ast}}{A_{0}}\Delta_{0}^{\ast}\Delta_{2}^{\ast}\right) 
+\frac{1-|\epsilon_{K}|^{2}}{1 + |\epsilon_{K}|^{2}}Im\left(\frac{A_{2}^{\ast}}{A_{0}}
\Delta_{0}^{\ast}\Delta_{2}^{\ast}\right)]\cos(\delta_{0}-\delta_{2}) \nonumber 
\end{eqnarray}
where we have neglected quadratic terms of $\omega=|A_{2}/A_{0}|$. Note that the above results 
hold for any choice of phase conventions. It is then obvious that
\begin{equation}
a_{\varepsilon\Delta}^{0} = \frac{2}{3} \hat{a}_{\epsilon'}^{(+-)} + 
\frac{1}{3} \hat{a}_{\epsilon'}^{(00)}
\end{equation}
which shows that once the asymmetries $\hat{a}_{\epsilon'}^{(+-)}$ and 
$\hat{a}_{\epsilon'}^{(00)}$ are measured, their combination given above will 
allow one to extract a clean signature of CPT violation arising from the decay amplitudes.
Where the values of $\hat{a}_{\epsilon'}^{(+-)}$ and 
$\hat{a}_{\epsilon'}^{(00)}$  can be simply extracted from the asymmetry $A_{CP+CPT}(t)$ at $t=0$
in eq. (39). It is noticed that when $|\Delta_{0}|<< 1$, i.e., $|a_{\epsilon +\epsilon'_{\Delta\Delta}}^{0}|
<< |a_{\epsilon + \epsilon'}^{0}|$ (while $\Delta_{2}$ could remain at the order of one), one has 
\begin{equation}
a_{\epsilon + \epsilon'}^{0} \simeq \frac{2}{3} \hat{a}_{\epsilon + \epsilon'}^{(+-)} + 
\frac{1}{3} \hat{a}_{\epsilon + \epsilon'}^{(00)}
\end{equation}
which indicates that by measuring $\hat{a}_{\epsilon + \epsilon'}^{(+-)}$ and  
$\hat{a}_{\epsilon + \epsilon'}^{(00)}$ one may extract the direct-indirect 
mixed-induced CP violation.

\section{Conclusions}

   In summary, we have developed the general model-independent and rephase-invariant formalism for testing 
CP- and CPT-noninvariant observables in meson decays. The formalism presented in previous 
articles for CPT is based on the density matrix approach\cite{CPTV}. 
In our article, we present a complete time-dependent
and rephase-invariant formulation in terms of amplitudes. The rephase invariance of all CP and CPT 
noninvariant observables is maintained throughout the calculation. 
All possible independent observables have been
classified systematically, which is more general and complete than the published results and can be used
for all meson decays. This enables one to separately measure different types of CPT- and CP-violating observables and 
to neatly distinguish effects of CPT from CP violation. The formalism which involves
many and elaborate definitions is directly related to fundamental parameters and 
can prove advantageous in establishing CPT-violating parameters from time-dependent 
measurements of meson decays. Several time-dependent CPT- and CP- asymmetries have been 
introduced, which led to some interesting observations: 

i). {\it As long as measurements of the asymmetry $A_{CP+CPT}(t)$ in the neutral meson 
decays (classified in the scenario i) in section 5 ) is not 
a constant but depends on time, one can conclude that CPT invariance is broken due to mixing}; 

ii). For the semileptonic decays $M^{0} \rightarrow M^{'-} l\nu$, one may come to a strong statement that   
{\it once the asymmetry $A_{CP+CPT}(t)$ is not zero, then CPT must be violated}. Among the decays the 
semileptonic decays are the more representative and perhaps the easiest to measure.

iii). A combined measurement of several time-dependent CPT- and CP- asymmetries from semileptonic and 
nonleptonic decays is necessary in order to isolate separately the indirect 
and direct CPT- and CP-violating effects.
 
Extraction of a clean signature on CPT, CP and T violation will play an important role 
in testing the standard model and local quantum field theory and in addition  
provides an interesting window for probing new physics. For all these reasons, this topic attracts 
a lot of attention\cite{CPTR}. We hope that the general rephase-invariant formalism presented 
in this paper will be useful for further studies of CPT, CP and T in the 
neutral meson systems produced at B-factories, the $\Phi$-factory\cite{DAFNA} and colliders.
 
 {\bf Acknowledgements:}  W.P was supported in part by the US department of Energy, Division of High 
Energy Physics, under Grant DOE/ER/0 1545-778. Two of us (E.A.P and Y.L.W) thanks  
Bundesministerium f\"{u}r Bildung, Wissenschaft, Forschung und Technologie (BMBF), 057D093P(7), Bonn, 
FRG, and DFG Antrag PA-10-1 for the financial support. Y.L.W acknowledges the support 
 by the NSF of China under Grant 19625514.

\newpage

{\bf Appendix }
\begin{appendix}

Here we collect some useful formuli.

 The definitions for the rephase-invariant observables: 
\begin{eqnarray}
\hat{a}_{\epsilon'} & = & \frac{1 - |h/g|^2}{1 + |h/g|^2} = 
\frac{2 Re \varepsilon_{M}'}{1 + |\varepsilon_{M}'|^2} \ , \nonumber \\
\hat{a}_{\epsilon_{S} + \epsilon'} & = & 
\frac{-4 Im(q_{S}h/p_{S}g)}{(1+|q_{S}/p_{S}|^2)(1+|h/g|^2)} \nonumber \\
& = & \frac{1}{1-a_{\epsilon}a_{\Delta}} 
\left[\hat{a}_{\epsilon + \epsilon'} \sqrt{1-a_{\Delta}^{2}-a_{\Delta}^{'2} } 
-a'_{\Delta} (1 + \hat{a}_{\epsilon \epsilon'} ) \right]\ , \hspace{1.2cm} [A1] \nonumber \\
\hat{a}_{\epsilon_{S} \epsilon'} & = & 
\frac{4 Re(q_{S}h/p_{S}g)}{(1+|q_{S}/p_{S}|^2)(1+|h/g|^2)} -1  \nonumber \\
& = &  \frac{1}{1-a_{\epsilon}a_{\Delta}}\left[ \hat{a}_{\epsilon\epsilon'}
\sqrt{1-a_{\Delta}^{2}-a_{\Delta}^{'2} } 
 + a'_{\Delta} \hat{a}_{\epsilon +\epsilon'}  
+(\sqrt{1-a_{\Delta}^{2}-a_{\Delta}^{'2} } - 1) 
+ a_{\epsilon}a_{\Delta} \right]\ ,   \nonumber 
\end{eqnarray}
with
\begin{eqnarray}
\hat{a}_{\epsilon + \epsilon'} & = & \frac{-4 Im(qh/pg)}{(1+|q/p|^2)(1+|h/g|^2)} 
=\frac{2 Im\epsilon_{M}(1-|\varepsilon_{M}'|^2) + 2 Im\varepsilon_{M}'
(1-|\epsilon_{M}|^2) }{ (1 + |\epsilon_{M}|^2 )(1+|\varepsilon_{M}'|^2)}\ ,  \nonumber \\
\hat{a}_{\epsilon\epsilon'} & = & \frac{4 Re(qh/pg)}{(1+|q/p|^2)(1+|h/g|^2)} -1 
= \frac{4 Im\epsilon_{M} \  Im \varepsilon_{M}' - 2
(|\epsilon_{M}|^2 + |\varepsilon_{M}'|^2) }{( 1 + |\epsilon_{M}|^2)
( 1 + |\varepsilon_{M}'|^2)}. \hspace{0.3cm} [A2] \nonumber
\end{eqnarray}

Rephase invariant observables for purely CP and CPT violation
\begin{eqnarray}
a_{\epsilon''} & = & \frac{|\sum_{i} A_{i} e^{i\delta_{i}}|^{2} -
 |\sum_{i} A_{i}^{\ast} e^{i\delta_{i}}|^{2} }{|\sum_{i} A_{i} e^{i\delta_{i}}|^{2} 
+ |\sum_{i} A_{i}^{\ast} e^{i\delta_{i}}|^{2} } =-\frac{2\sum_{ij}A_{i}A_{j}^{\ast}
\sin(\delta_{i}-\delta_{j})}{|\sum_{i} A_{i} e^{i\delta_{i}}|^{2} 
+ |\sum_{i} A_{i}^{\ast} e^{i\delta_{i}}|^{2} } \  ,
\nonumber \\
a_{\varepsilon\Delta} & = & \frac{2\sum_{i,j} A_{i}A_{j}^{\ast}(\Delta_{i} 
+ \Delta_{j}^{\ast}) \cos (\delta_{i} - \delta_{j})}{ 
|\sum_{i} A_{i} e^{i\delta_{i}}|^{2} + |\sum_{i} A_{i}^{\ast} e^{i\delta_{i}}|^{2} }, 
\nonumber \\
a'_{\varepsilon\Delta} & = & \frac{2i\sum_{i,j} A_{i}A_{j}^{\ast}(\Delta_{i} 
+ \Delta_{j}^{\ast}) \sin (\delta_{i} - \delta_{j})}{ 
|\sum_{i} A_{i} e^{i\delta_{i}}|^{2} + |\sum_{i} A_{i}^{\ast} e^{i\delta_{i}}|^{2} }, 
\hspace{1.8cm} [A3] \nonumber \\
a_{\Delta\Delta} & = & \frac{2\sum_{i,j} A_{i}A_{j}^{\ast}\Delta_{i} 
\Delta_{j}^{\ast} \cos (\delta_{i} - \delta_{j}}{ 
|\sum_{i} A_{i} e^{i\delta_{i}}|^{2} + |\sum_{i} A_{i}^{\ast} e^{i\delta_{i}}|^{2} },
 \nonumber  \\
a'_{\Delta\Delta} & = & \frac{2i\sum_{i,j} A_{i}A_{j}^{\ast}\Delta_{i} 
\Delta_{j}^{\ast} \sin (\delta_{i} - \delta_{j})}{ 
|\sum_{i} A_{i} e^{i\delta_{i}}|^{2} + |\sum_{i} A_{i}^{\ast} e^{i\delta_{i}}|^{2} }
\nonumber 
\end{eqnarray}
with $\Delta_{i} = B_{i}/A_{i}$. Here $\Delta_{i}$ are rephase-invariant quantities 
and characterize direct CPT violation in the decay amplitudes. 

\begin{eqnarray}
a_{\epsilon + \epsilon'} & = & \frac{2 Im\epsilon_{M}(1-|\epsilon_{M}'|^2) 
+ 2 Im\epsilon'_{M}(1-|\epsilon_{M}|^2) }{ (1 + |\epsilon_{M}|^2 )(1+|\epsilon_{M}'|^2)} 
\nonumber \\
a_{\epsilon + \epsilon'_{\Delta}} & = & \frac{2 Im\epsilon_{M}(1-|\epsilon'_{\Delta}|^2) 
+ 2 Im\epsilon'_{\Delta}(1-|\epsilon_{M}|^2) }{ 
(1 + |\epsilon_{M}|^2 )(1+|\epsilon'_{\Delta}|^2)}\ , \hspace{1.0cm} [A4] \nonumber \\ 
a_{\epsilon + \epsilon'_{\Delta\Delta}} & = & 
\frac{2 Im\epsilon_{M}(1-|\epsilon'_{\Delta\Delta}|^2) 
+ 2 Im\epsilon'_{\Delta\Delta}(1-|\epsilon_{M}|^2) }{ 
(1 + |\epsilon_{M}|^2 )(1+|\epsilon'_{\Delta\Delta}|^2)}\  , \nonumber 
\end{eqnarray}
with 
\begin{eqnarray}
\frac{1- |\epsilon'_{M}|^{2}}{1 + |\epsilon'_{M}|^{2}} & = & 
\frac{2\sum_{i,j}Re (A_{i}A_{j}) \cos(\delta_{i} - \delta_{j}) }{|\sum_{i} 
A_{i} e^{i\delta_{i}}|^{2} + |\sum_{i} A_{i}^{\ast} e^{i\delta_{i}}|^{2}} \  ,
\nonumber \\
\frac{1- |\epsilon'_{\Delta}|^{2}}{1 + |\epsilon'_{\Delta}|^{2}} & = & 
-\frac{2\sum_{i,j} Im [A_{i}A_{j}(\Delta_{i} - \Delta_{j})] 
\sin(\delta_{i} - \delta_{j}) }{|\sum_{i} A_{i} e^{i\delta_{i}}|^{2} 
+ |\sum_{i} A_{i}^{\ast} e^{i\delta_{i}}|^{2}} \  ,\nonumber \\
\frac{1- |\epsilon'_{\Delta\Delta}|^{2}}{1 + |\epsilon'_{\Delta\Delta}|^{2}} & = & 
-\frac{2\sum_{i,j} Re[A_{i}A_{j}(\Delta_{i} \Delta_{j})] 
\cos(\delta_{i} - \delta_{j}) }{|\sum_{i} A_{i} e^{i\delta_{i}}|^{2} 
+ |\sum_{i} A_{i}^{\ast} e^{i\delta_{i}}|^{2}}\  , \nonumber \\
\frac{2Im \epsilon'_{M}}{1 + |\epsilon'_{M}|^{2}} & = & 
-\frac{2\sum_{i,j}Im (A_{i}A_{j}) \cos(\delta_{i} - \delta_{j}) }{|\sum_{i} 
A_{i} e^{i\delta_{i}}|^{2} + |\sum_{i} A_{i}^{\ast} e^{i\delta_{i}}|^{2}}\ , 
\hspace{1.8cm} [A5] \nonumber \\
\frac{2Im\epsilon'_{\Delta}}{1 + |\epsilon'_{\Delta}|^{2}} & = & 
-\frac{2\sum_{i,j} Re[A_{i}A_{j}(\Delta_{i} - \Delta_{j})] 
\sin(\delta_{i} - \delta_{j}) }{|\sum_{i} A_{i} e^{i\delta_{i}}|^{2} 
+ |\sum_{i} A_{i}^{\ast} e^{i\delta_{i}}|^{2}}\  , \nonumber \\
\frac{2Im\epsilon'_{\Delta\Delta}}{1 + |\epsilon'_{\Delta\Delta}|^{2}} & = & 
\frac{2\sum_{i,j} Im[A_{i}A_{j}(\Delta_{i} \Delta_{j})] 
\cos(\delta_{i} - \delta_{j}) }{|\sum_{i} A_{i} e^{i\delta_{i}}|^{2} 
+ |\sum_{i} A_{i}^{\ast} e^{i\delta_{i}}|^{2}}\ . \nonumber 
\end{eqnarray}

The exact expressions for the time-dependent CP and CPT asymmetries in the scenario i): 

\begin{eqnarray}
A_{CP+CPT}(t) & = & \frac{\Gamma(M^{0}(t) \rightarrow f) - \Gamma 
(\overline{M}^{0} (t) \rightarrow \bar{f}) }{\Gamma(M^{0}(t) \rightarrow f) + \Gamma 
(\overline{M}^{0} (t) \rightarrow \bar{f} )}\  ,  \nonumber \\
& = & \frac{\hat{a}_{\epsilon''} + 2{\cal A}_{CPT}(t)/[ (1+|\eta_{\Delta}|^{2})
\cosh \Delta\Gamma t +(1-|\eta_{\Delta}|^{2})\cos \Delta m t ]}{ 1 + 
2\hat{a}_{\epsilon''} {\cal A}_{CPT}(t)/ [ (1+|\eta_{\Delta}|^{2})
\cosh \Delta\Gamma t +(1-|\eta_{\Delta}|^{2})\cos \Delta m t ]}\ , \hspace{0.1cm} [A6] \nonumber  \\
A'_{CP+CPT} (t) & = & \frac{\Gamma(\overline{M}^{0}(t) \rightarrow f) - 
\Gamma (M^{0}(t) \rightarrow \bar{f} )}{\Gamma(\overline{M}^{0}(t) 
\rightarrow f) + \Gamma (M^{0} (t) \rightarrow \bar{f} )}\  , \nonumber \\
& = & (\hat{a}_{\epsilon''} + \frac{2 a_{\epsilon}}{1 + a_{\epsilon}^{2}})/(1 +
\frac{2 a_{\epsilon}}{1 + a_{\epsilon}^{2}}\hat{a}_{\epsilon''})\  , \hspace{2.5cm} [A7] \nonumber \\
A''_{CP+CPT}(t) & = & \frac{\Gamma(M^{0}(t) \rightarrow f) - \Gamma (\overline{M}^{0}(t)
\rightarrow f )}{\Gamma(M^{0}(t) \rightarrow f) + 
\Gamma(\overline{M}^{0} (t) \rightarrow f )}\  , \nonumber \\
& = & \frac{\frac{1-a_{\Delta}^{2}-a_{\Delta}^{'2}
+ a_{\epsilon}a_{\Delta}^{'2}}{(1-a_{\epsilon})(1-a_{\Delta}^{'2})}\cos \Delta m t 
- \frac{a_{\epsilon} - a_{\Delta}^{2}}{(1-a_{\epsilon})(1-a_{\Delta}^{'2})} 
\cosh \Delta \Gamma t + {\cal A}_{CPT}(t) }{\frac{1-a_{\epsilon} a_{\Delta}^{2}}{
(1-a_{\epsilon})(1-a_{\Delta}^{'2})} \cosh \Delta \Gamma t 
- \frac{a_{\epsilon}(1-a_{\Delta}^{2}-a_{\Delta}^{'2}) + a_{\Delta}^{'2}}{
(1-a_{\epsilon})(1-a_{\Delta}^{'2})} \cos \Delta m t + {\cal A}_{CPT}(t) }  \  , 
\hspace{0.3cm} [A8] \nonumber
\end{eqnarray}
with 
\[
{\cal A}_{CPT} =  -\frac{a_{\Delta}}{1-a_{\Delta}^{'2}} \sinh \Delta \Gamma t
+ \frac{a'_{\Delta}\sqrt{1- a_{\Delta}^{2} - a_{\Delta}^{'2}}}{1-a_{\Delta}^{'2}}
\sin \Delta m t \  , \hspace{0.8cm} [A9] 
\]
and in the scenario ii):

\[
%\begin{equation}
A_{CP+CPT}(t)  =  \frac{\hat{\Delta}_{m}(t) + \Delta_{CPT}(t) - a_{\epsilon_{S}}\ 
(\hat{\Delta}_{\gamma}(t) + \Delta'_{CPT}(t))}{\hat{\Delta}_{\gamma}(t) 
+ \Delta'_{CPT}(t) - a_{\epsilon_{S}}(\hat{\Delta}_{m}(t) + \Delta_{CPT}(t))}\ , \hspace{0.8cm} [A10] 
%\end{equation}
\]
with
\begin{eqnarray} 
& & \hat{\Delta}_{m}(t) = (a_{\epsilon_{S}} + \hat{a}_{\epsilon'} ) \cos (\Delta m t) + 
\hat{a}_{\epsilon_{S} + \epsilon'}   \sin (\Delta m t)  \nonumber \\
& & \hat{\Delta}_{\gamma}(t) = (1 + a_{\epsilon_{S}} \hat{a}_{\epsilon'} ) 
\cosh (\Delta \Gamma t)  + (1 + \hat{a}_{\epsilon_{S} \epsilon'} )
 \sinh (\Delta \Gamma t) \  , \hspace{0.8cm} [A11] \nonumber   
\end{eqnarray}
and 
\begin{eqnarray}
\Delta_{CPT}(t) & = & (2+a_{\epsilon_{S}}\hat{a}_{\epsilon'} + 
\hat{a}_{\epsilon_{S}\epsilon'}) [\frac{a_{\Delta}}{1-a_{\Delta}^{'2}}
(\cos\Delta m t - e^{\Delta\Gamma t}) + \frac{a'_{\Delta}}{1-a_{\Delta}^{'2}}
\sin\Delta m t ] \nonumber \\
\Delta'_{CPT}(t) & = &  -[\frac{a_{\Delta}}{1-a_{\Delta}^{'2}}
(a_{\epsilon_{S}} + \hat{a}_{\epsilon'} ) + \frac{a'_{\Delta}}{1-a_{\Delta}^{'2}}
\hat{a}_{\epsilon_{S} + \epsilon'} ](\cos\Delta m t - e^{-\Delta\Gamma t}) \nonumber \\
& + & [\frac{a'_{\Delta}}{1-a_{\Delta}^{'2}}
(a_{\epsilon_{S}} + \hat{a}_{\epsilon'} ) - \frac{a_{\Delta}}{1-a_{\Delta}^{'2}}
\hat{a}_{\epsilon_{S} + \epsilon'} ]\sin\Delta m t \  , \hspace{1.5cm} [A12] \nonumber \\
& - & (2+a_{\epsilon_{S}}\hat{a}_{\epsilon'} + 
\hat{a}_{\epsilon_{S}\epsilon'}) \frac{a_{\Delta}^{2} + 
a_{\Delta}^{'2} }{(1-a_{\Delta}^{'2})^{2}} (\cos\Delta m t - \cosh\Delta\Gamma t) \nonumber 
\end{eqnarray}
Note that when CPT is conserved, $\Delta_{CPT}(t) = \Delta'_{CPT} = 0$. 
\end{appendix}

\newpage

%\begin{references}

%\end{references}

\end{document}